\documentclass[journal]{IEEEtran}
\usepackage{amsthm,amsmath,comment,bbm,amssymb,color,graphicx}
\usepackage{mathtools}
\usepackage{comment}
\usepackage{url}
\usepackage{caption}
\usepackage{subcaption}
\usepackage{epstopdf}
\usepackage{tikz}
\usetikzlibrary{positioning,shapes,shadings,backgrounds,calc}
\usepackage{multirow}
\usepackage{multicol}
\def\nb0{{\mathbf{0}}}
\def\nb1{{\mathbf{1}}}

\newtheorem{nrem}{Remark}

\newtheorem{eg}{Example}

\begin{document}
\title{Rate-Splitting Multiple Access for 6G: Prototypes, Experimental Results and Link/System level Simulations}
\author{Sundar Aditya, Yong Jin Daniel Kim, David Vargas, David Redgate, Onur Dizdar, Neil Bhushan, Xinze Lyu, Sibo Zhang, Stephen Wang and Bruno Clerckx
\thanks{S. Aditya, X. Lyu and B. Clerckx are with the Dept.~of Electrical and Electronic Engg., Imperial College London, London SW7 2AZ, UK. Their work was funded by UKRI Impact Acceleration Account (IAA) grant EP/X52556X/1, and EPSRC grants EP/X040569/1, EP/Y037197/1, EP/X04047X/1, EP/Y037243/1.}
\thanks{Y. J. D. Kim is with the Department of Electrical and Computer Engg., Rose-Hulman Institute of Technology, Terre Haute, IN 47803, USA.}
\thanks{D. Vargas is with BBC Research \& Development, London W12 7TQ, UK.}
\thanks{D. Redgate, O. Dizdar, N. Bhushan and S. Wang are with VIAVI Marconi Labs, VIAVI Inc., Stevenage SG1 2AN, UK.}
\thanks{S. Zhang is with the Dept.~of Electrical and Electronic Engg., Imperial College London, London SW7 2AZ, UK, and also with BBC Research \& Development, London W12 7TQ, UK. His work was supported by EPSRC Industrial Case Award no. 210163.}
}
\maketitle

\begin{abstract}
Rate-Splitting Multiple Access (RSMA) is a powerful and versatile physical layer multiple access technique that generalizes and has better interference management capabilities than 5G-based Space Division Multiple Access (SDMA). It is also a rapidly maturing technology, all of which makes it a natural successor to SDMA in 6G. In this article, we describe RSMA's suitability for 6G by presenting: i) link and system level simulations of RSMA's performance gains over SDMA in realistic environments, and (ii) pioneering experimental results that demonstrate RSMA's gains over SDMA for key use cases like enhanced Mobile Broadband (eMBb), and Integrated Sensing and Communications (ISAC). We also comment on the status of standardization activities for RSMA.
\end{abstract}

\begin{IEEEkeywords}
Rate-Splitting Multiple Access, RSMA for 6G, RSMA prototyping, RSMA for eMBb, RSMA for ISAC, RSMA in 3GPP
\end{IEEEkeywords}

\bstctlcite{IEEEexample:BSTcontrol}

\section{Introduction}
As with previous evolutions, 6G will seek to realize enhancements over 5G in terms of scale (number of simultaneous users), performance (quality of service (QoS)) and versatility (support for new use cases, such as integrated sensing and communications (ISAC)). The interference management capabilities of the underlying physical layer multiple access technique are central to realizing these targets. The technique used in multiple-input multiple-output (MIMO) based 5G is Space Division Multiple Access (SDMA), enabled by linear precoding.

Rate-Splitting Multiple Access (RSMA) is a powerful and versatile MIMO-based physical layer multiple access technique that is well-suited to meet the ambitious targets of 6G: \emph{powerful} because its interference management capabilities are superior to SDMA and Non-Orthogonal Multiple Access (NOMA, an alternative to SDMA first considered for the uplink in 5G); and \emph{versatile} because it can provide better QoS for several use cases, including many emerging ones \cite{RSMA_JSAC_Primer}. Importantly, RSMA generalizes both SDMA and NOMA, i.e., under favourable channel conditions for SDMA/NOMA, RSMA automatically reduces to these schemes, but achieves strictly better performance than both SDMA and NOMA for (the vast majority of) channel conditions where neither scheme is the most effective. These attractive features make RSMA a natural successor to SDMA in 6G.  

While the above benefits of RSMA have been well-documented in theory \cite{RSMA_JSAC_Primer, mao2022fundmental}, the path to standardization involves demonstrating RSMA's gains in realistic deployments, which is the focus of this article. We begin in Section~\ref{sec:rsma_overview} by providing an overview of RSMA, shedding light on its powerful interference management strategy. Then, in Section~\ref{sec:rsma_implementation}, we consider RSMA's suitability for 6G as follows:
\begin{itemize}
    \item In Section~\ref{subsec:lls}, we present link level simulations, where the Physical Downlink Shared Channel (PDSCH) in the 5G New Radio (NR) standards is adapted to implement RSMA. For the enhanced Mobile Broadband (eMBb) use case in particular, we observe that RSMA can yield a 3dB SNR gain in the block error rate (BLER) over SDMA.

    \item Next, in Section~\ref{subsec:sls}, we present system level simulations of RSMA's eMBb performance in indoor hotspot and urban microcell environments. We show that in such realistic environments, RSMA achieves greater fairness at higher sum rates than SDMA.
        
    \item Finally, in Section~\ref{subsec:prototypes}, we present experimental results -- obtained from software-defined radio based prototypes -- for key use cases like  eMBb and ISAC. For eMBb in particular, RSMA achieves greater fairness at higher sum rates than both SDMA and NOMA.
\end{itemize}
We then conclude this article with a few comments on the status of RSMA standardization activities.

\section{RSMA Overview}
\label{sec:rsma_overview}
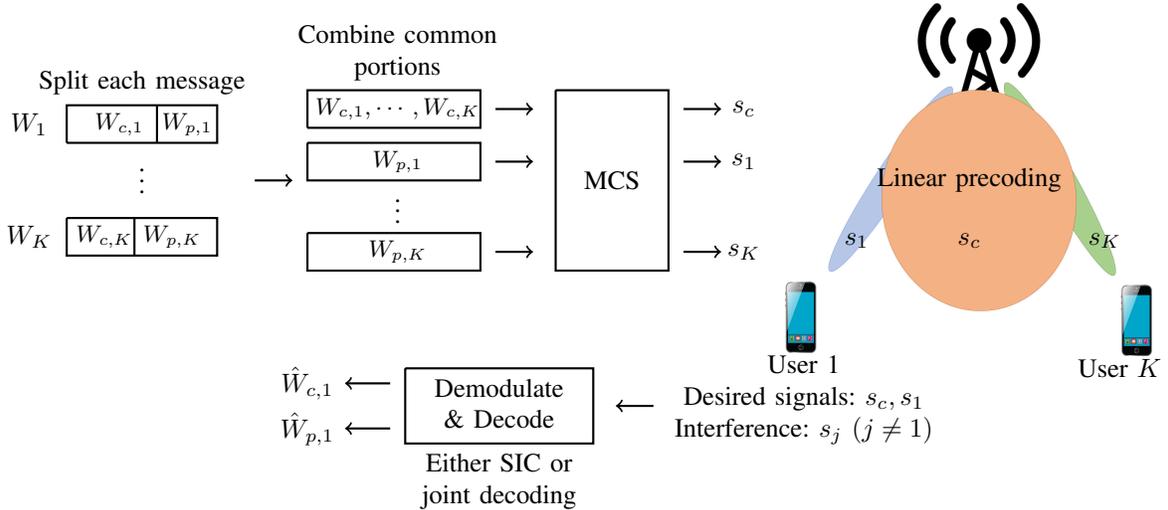
\begin{figure*}
    \centering
    \begin{tikzpicture}
        \draw[thick,-] (0,0) -- (2,0) -- (2,0.5) -- (0,0.5) -- (0,0) node at (-0.5,0.25) {$W_1$} node at (0.7,0.25){\small $W_{c,1}$} node at (1.6,0.25) {\small $W_{p,1}$};
        \draw[thick,-] (1.2,0) -- (1.2,0.5);
        \node at (1,-0.4) {$\vdots$};
        \draw[thick,-] (0,-1) -- (2,-1) -- (2,-1.5) -- (0,-1.5) -- (0,-1) node at (-0.5,-1.25) {$W_K$} node at (0.5,-1.25){\small $W_{c,K}$} node at (1.4,-1.25) {\small $W_{p,K}$};
        \draw[thick,-] (0.9,-1) -- (0.9,-1.5);
        \node at (1,0.8) {Split each message};
        \draw[thick,->] (2.5,-0.5) -- (3,-0.5);
        \draw[thick,-] (3.2,0.2) -- (5.5,0.2) -- (5.5,0.7) -- (3.2,0.7) -- (3.2,0.2) node at (4.4,0.45) {\small $W_{c,1}, \cdots, W_{c,K}$} node at (4.4, 1.2){\begin{tabular}{c}
            Combine common  \\
             portions
        \end{tabular}};
        \draw[thick,-] (3.2,0) -- (5.5,0) -- (5.5,-0.5) -- (3.2,-0.5) -- (3.2,0) node at (4.4,-0.25) {\small $W_{p,1}$} node at (4.4, -0.8){$\vdots$};
        \draw[thick,-] (3.2,-1.2) -- (5.5,-1.2) -- (5.5,-1.7) -- (3.2,-1.7) -- (3.2,-1.2) node at (4.4,-1.45) {\small $W_{p,K}$};
        \draw[thick,->] (5.7,0.45) -- (6.2,0.45);
        \draw[thick,->] (5.7,-0.25) -- (6.2,-0.25);
        \draw[thick,->] (5.7,-1.45) -- (6.2,-1.45);
        \draw[thick,-] (6.5,0.7) -- (8,0.7) -- (8,-1.7) -- (6.5,-1.7) -- (6.5,0.7) node at (7.25,-0.5){MCS};
        \draw[thick,->] (8.2,0.45) -- (8.7,0.45) node at (9,0.45){$s_c$};
        \draw[thick,->] (8.2,-0.25) -- (8.7,-0.25) node at (9,-0.25){$s_1$};
        \draw[thick,->] (8.2,-1.45) -- (8.7,-1.45) node at (9,-1.45){$s_K$};
        \node[inner sep=0pt] (BS) at (12,-0.5)
    {\includegraphics[scale=0.15]{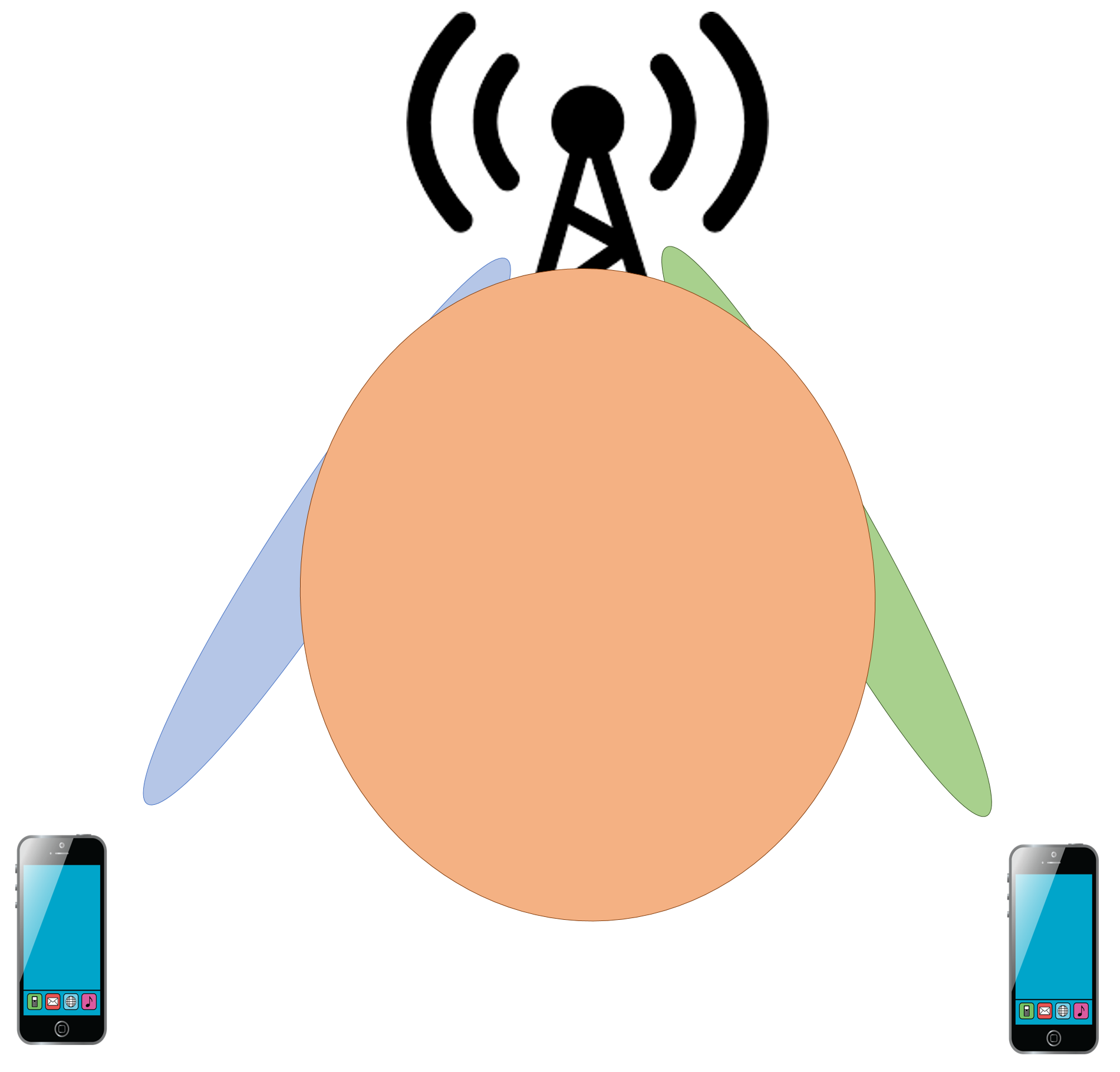}};
        \node at (12,-1.3){$s_c$};
        \node at (10.5,-1.3){$s_1$};
        \node at (13.8,-1.3){$s_K$};
        \node at (12,-0.5){Linear precoding};
        \node at (9.8,-3.4){\begin{tabular}{c}
            User 1 \\
          Desired signals: $s_c, s_1$ \\
          Interference: $s_j ~(j\neq 1)$
        \end{tabular}};
        \node at (14,-3){User $K$};
        \draw[thick,->] (7.8,-3.5) -- (7.3,-3.5);
        \draw[thick,-] (4.5,-3) -- (7,-3) -- (7,-4) -- (4.5,-4) -- (4.5,-3) node at (5.75,-3.5){\begin{tabular}{c}
             Demodulate \\
             \& Decode 
        \end{tabular}};
        \node at (5.75,-4.5){\begin{tabular}{c}
           Either SIC or \\
           joint decoding
        \end{tabular}};
        \draw[thick,->] (4.3,-3.2) -- (3.7,-3.2) node at (3.2,-3.2){$\hat{W}_{c,1}$};
        \draw[thick,->] (4.3,-3.8) -- (3.7,-3.8) node at (3.2,-3.8){$\hat{W}_{p,1}$};
    \end{tikzpicture}
    \caption{Illustration of downlink RSMA. The $i$-th user's message $W_i = [W_{c,i}, W_{p,i}]$ and the sizes of $W_{c,i}$ and $W_{p,i}$ can vary across $i$, in general. Since $s_c$ needs to be decoded by all users, its precoder must have sufficient gain over all $K$ user channels. In constrast, since $s_i$ needs to be decoded by only user $i$, its precoder must have high gain over user $i$'s channel to maximize the desired signal's strength, and low gain over all other user channels to minimize the interference.}
    \label{fig:RSMA_principle}
\end{figure*}
To illustrate RSMA in operation, consider the downlink in Fig.~\ref{fig:RSMA_principle}, where a $N_T$-antenna base station (BS) communicates with $K$ users, intending to convey message $W_i$ to user~$i \in \{1, \cdots, K\}$. At the BS, each $W_i$ is split into two parts – $W_{c,i}$ (known as the common portion) and $W_{p,i}$ (known as the private portion). The common portions -- $W_{c,1}, \cdots, W_{c,K}$ -- are aggregated to form a common message $W_c$, which is then encoded and modulated (through suitable choice of modulation and coding scheme (MCS) level) to form a data stream, $s_c$, known as the \emph{common stream}. Similarly, each $W_{p,i}$ is individually encoded and modulated to form a data stream, $s_i$, known as the $i$-th \emph{private stream}. The $K+1$ data streams are then individually precoded (through appropriate choices from a precoding codebook, for instance), and transmitted over the air.

\begin{nrem}
\label{rem:extra_stream}
    The key feature of RSMA is the transmission of $K+1$ precoded data streams to convey $K$ messages to $K$ users simultaneously\footnote{Strictly speaking, this refers to one-layer RSMA. For a complete description of all RSMA variants, see \cite{RSMA_JSAC_Primer}.}. In contrast, SDMA and NOMA both use $K$ precoded streams for this purpose. The extra precoded stream is more effective at adapting to channel conditions to suppress interference, which in turn improves spectral efficiency and fairness. The extra precoded stream also yields performance gains for new applications, such as ISAC. 
\end{nrem}

At user~$i$, the received signal is a mixture of the $K+1$ streams, with different strengths determined by the BS $\rightarrow$ user~$i$ channel and the choice of precoders. To retrieve the desired message $W_i$, user~$i$ needs to decode two streams, namely $s_c$ and $s_i$, as $W_i$ is split between them. The two streams can either be decoded jointly (known as joint decoding) or successively, where $s_c$ is decoded first followed by $s_i$ (this is known as successive interference cancellation (SIC)). The latter approach increases the receiver complexity and comes with the risk of error propagation when $s_c$ is incorrectly decoded. To address this frequently encountered criticism of RSMA, \cite{SiboTCOM} investigated several low-complexity non-SIC receiver architectures that avoid canceling the common stream before decoding private streams. We discuss their performance in Section~\ref{subsec:lls}.

To understand the effectiveness of RSMA, one can broadly conceive two distinct strategies to managing interference at the users:
\begin{itemize}
    \item[a)] \emph{Fully Decoding Interference}, wherein a user decodes not just its desired message, but the messages of all other users with weaker channels than itself. This is the strategy adopted by NOMA, which is most effective when the mutual interference between all pairs of links is high (specifically, much higher than the noise levels).

    \item[b)] \emph{Treating Interference as Noise}, wherein each user decodes only its desired message by assuming the interference to have the same statistical characteristics as noise. This strategy is most effective when the mutual interference between all pairs of links is low (specifically, much lower than the noise levels). This is also the strategy adopted by SDMA, under the assumption that the choice of precoders sufficiently suppresses the signal leakage (i.e., interference) towards undesired users. 
\end{itemize}
Intuitively, for most sets of $K$ channels that do not fall under either extreme of high/low mutual interference between all pairs of links, the most effective interference management approach should be one that combines the above two strategies. RSMA achieves precisely this, whereby the interference is \emph{partially decoded, and partially treated as noise depending on channel conditions}. Specifically, while decoding $s_c$ and $s_i$ at user~$i$ (regardless of whether joint decoding or SIC is used), some of the interference caused by $W_j~(j \neq i)$ is decoded -- namely the common portion $W_{c,j}$ -- while the rest of the interference from $W_{p,j}$ (captured by $s_j$) is treated as noise.  

\begin{nrem}
\label{rem:rsma_general}
When $W_i = W_{p,i}$ for all $i$ (i.e., there is no common message $W_c$), RSMA reduces to SDMA. For the two-user case, suppose user 1 has a stronger channel than user 2. Then, for $W_1 = W_{p,1}$ and $W_2 = W_{c,2}$ (i.e., no common portion for the strong user and no private portion for the weak user), RSMA reduces to NOMA. Thus, SDMA and NOMA are special cases of RSMA corresponding to specific choices of message splitting\footnote{For $K> 2$, NOMA is a special case of generalized-RSMA, a variant of RSMA described in \cite{RSMA_JSAC_Primer}. See also footnote 1.}.     
\end{nrem}

From Remark~\ref{rem:rsma_general}, it follows that the performance gains of RSMA over SDMA/NOMA are the highest when the (i) splitting choice (i.e., how much of $W_i$ to allocate to common and private portions, $W_{c,i}$ and $W_{p,i}$, respectively), (ii) precoders and (iii) MCS levels for the $K+1$ streams are adapted to channel conditions. This can be realized through optimization (e.g., precoder design to maximize the sum rate) or link adaptation or a combination of both.

\section{RSMA in Realistic Deployments}
\label{sec:rsma_implementation}

\subsection{Link level Simulations}
\label{subsec:lls}
\begin{figure*}
    \centering
    \begin{tikzpicture}
    \node(a){\centering\includegraphics[width=\linewidth]{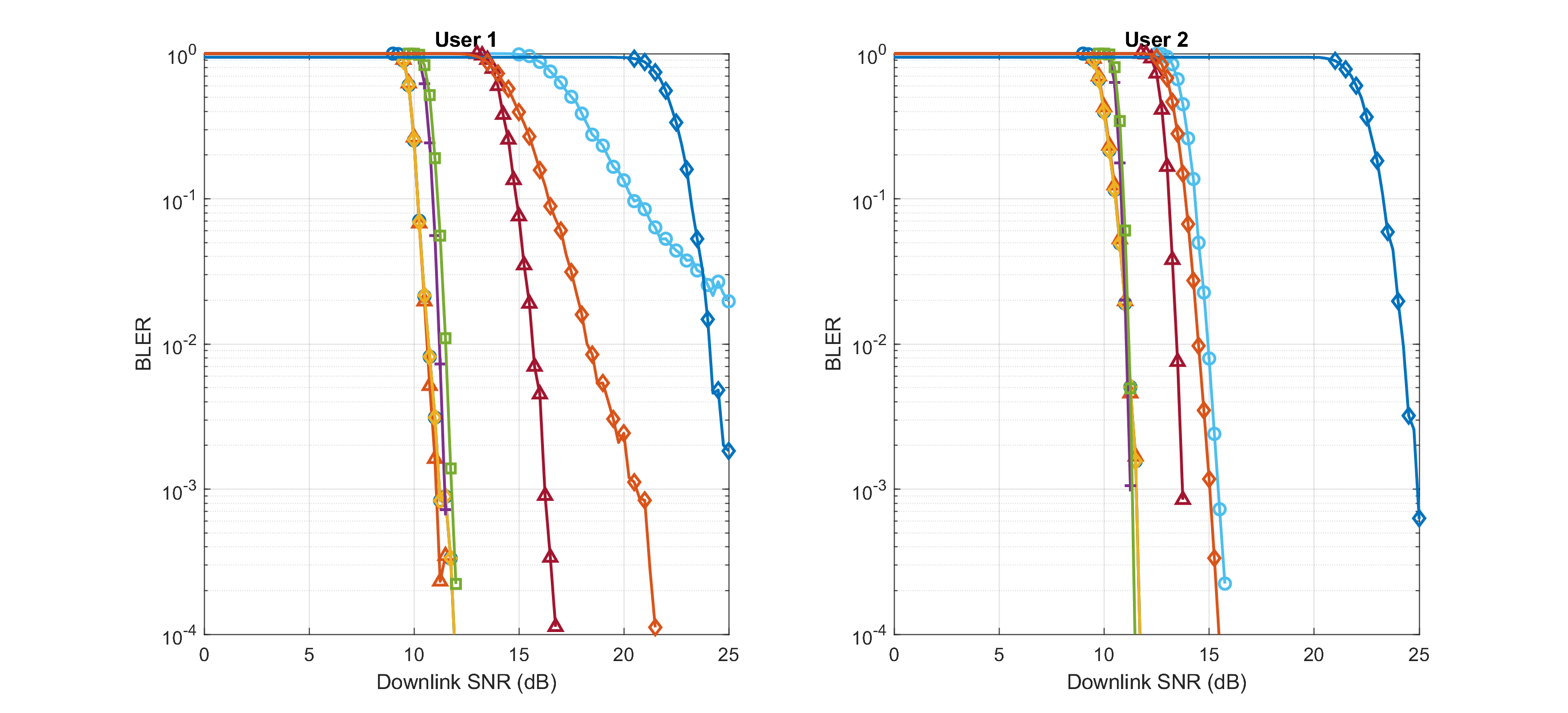}};
    \node(b)[draw, black,line width=1.2pt,ellipse, minimum width=30pt, minimum height=20pt,rotate=0,xshift=-115pt,yshift=-20pt]{};
    \node[above=0pt of b,xshift=-32pt,yshift=0pt]{\footnotesize RSMA};
    \node[above=0pt of b,xshift=-32pt,yshift=-8pt]{\footnotesize schemes};
    \node(c)[draw, black,line width=1.3pt,ellipse, minimum width=80pt, minimum height=30pt,rotate=0,xshift=-55pt,yshift=20pt]{};
    \node[right=-1pt of c,xshift=-50pt,yshift=-20pt]{\footnotesize SDMA};
    \node[right=-1pt of c,xshift=-45pt,yshift=-28pt]{\footnotesize schemes};

    \node(d)[draw, black,line width=1.2pt,ellipse, minimum width=25pt, minimum height=20pt,rotate=0,xshift=114pt,yshift=-20pt]{};
    \node[above=0pt of d,xshift=-32pt,yshift=0pt]{\footnotesize RSMA};
    \node[above=0pt of d,xshift=-32pt,yshift=-8pt]{\footnotesize schemes};
    
    \node(e)[draw, black,line width=1.3pt,ellipse, minimum width=90pt, minimum height=30pt,rotate=0,xshift=165pt,yshift=0pt]{};
    \node[right=-1pt of e,xshift=-60pt,yshift=-22pt]{\footnotesize SDMA};
    \node[right=-1pt of e,xshift=-55pt,yshift=-30pt]{\footnotesize schemes};     
    \end{tikzpicture} 
    \caption{Link level simulations: Downlink unicast (PDSCH) performance (transport block error rate versus SNR) of RSMA and SDMA at target spectral efficiencies of 3.4 bits per resource element. The two users are $20{\rm m}$ apart \cite{Sibo3GPP}.}
    \label{LLS_unicast}
\end{figure*}
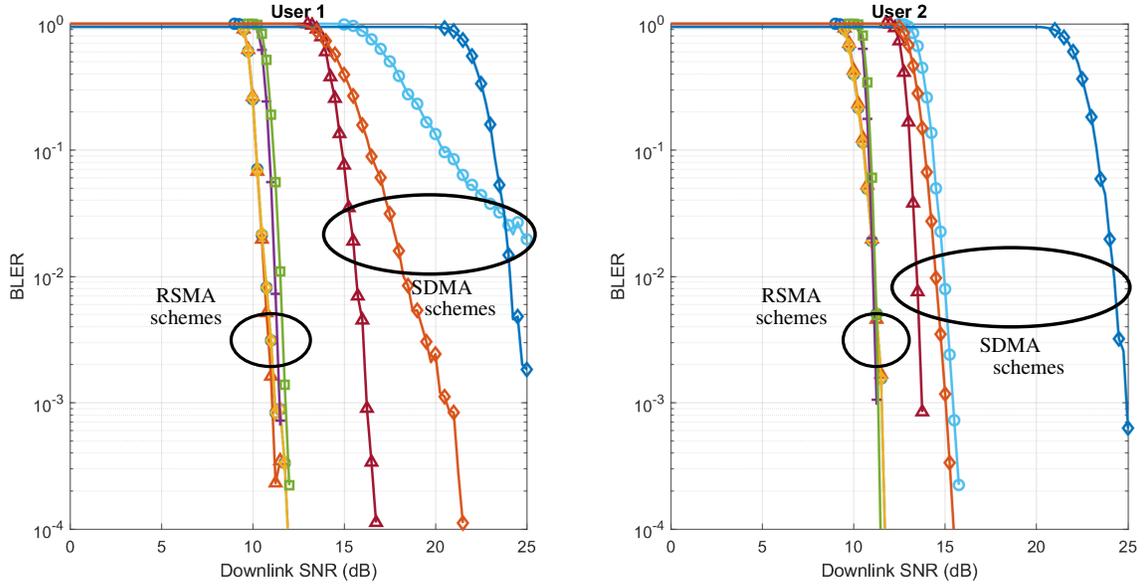

The canonical eMBb use case is downlink multi-user \emph{unicast} communications, where each user desires a unique message (as in Fig.~\ref{fig:RSMA_principle}). Fig.~\ref{LLS_unicast} presents link level simulations results for this use case involving two users $(K = 2)$ to get an insight into RSMA's gain over SDMA in realistic scenarios \cite{Sibo3GPP}. In these simulations, 5G NR's physical layer procedures for PDSCH such as waveform, frame structure, coding/decoding, uplink/downlink channel estimation, etc. were followed. Realistic channel modeling was obtained through ray tracing based on an approximate 3D geometry of the real-world environment. The different RSMA schemes in Fig.~\ref{LLS_unicast} correspond to different receiver architecture choices -- we consider SIC as well as all the non-SIC architectures in \cite{SiboTCOM}. The various SDMA schemes in Fig.~\ref{LLS_unicast} correspond to different precoder choices (see \cite{Sibo3GPP} for more details). We see that RSMA can achieve an SNR gain of around 3dB over SDMA. Importantly, the RSMA performance is largely the same for all receiver architectures. Thus, the low-complexity non-SIC receivers preserve most of RSMA's gains, while avoiding the error propagation in SIC. Hence, these alternative receiver architectures are promising candidates for implementation \cite{Sibo3GPP}.

It is well established in the literature (e.g. in \cite{SiboTCOM, RSMA_unicast_prototype, RSMA_NOUM_prototype}) that the gain from RSMA depends on the channel correlation among users. Since an important contributing factor to channel correlation is the distance between co-scheduled users, \cite{Sibo3GPP} also evaluated the performance of RSMA and SDMA for different inter-user separation distances. For users less than $20{\rm m}$ apart, RSMA provides significant gain (in Fig.~\ref{LLS_unicast}, the users are $20{\rm m}$ apart), but the gain is less noteworthy for distances greater than $40{\rm m}$.

\subsection{System level Simulations}
\label{subsec:sls}
Complementing the link level simulations, we simulate RSMA's system level performance for downlink unicast communications in two realistic network environments, namely indoor-office (InH) and urban-microcell (UMi) \cite{3gpp_release17}, as implemented in the QuaDRiGa simulator \cite{quadriga}. InH is intended to model typical offices as well as hotspot environments like shopping malls, while UMi models typical cities, urban squares, etc. Our simulation settings follow the full calibration parameters detailed in 3GPP TR 38.901 \cite[Table 7.8-2]{3gpp_release17}. In particular, we consider:
\begin{itemize}
    \item[a)] configuration 1 for BS antenna with 64 antenna elements mapped to 4 antenna ports, and one vertically polarized antenna with isotropic gain pattern for each user;
    \item[b)] $6{\rm GHz}$ carrier frequency, 50 resource blocks (RBs) with $400{\rm kHz}$ bandwidth per OFDM subcarrier.
\end{itemize}
The user densities (i.e., number of users per sq.~km) for InH and UMi have been selected based on 3GPP TS 22.261 \cite[Table 7.1-1]{3gpp_release20} to model \emph{High Demand Density} areas. For UMi, we assume fully outdoor users to model city square environments. We investigate the rate gains for RSMA over SDMA for $K$ randomly selected users (attached to the same sectors for the InH scenario, and attached to the south facing sector of the central site in the case of the UMi scenario). All results are compiled from 50,000 such independent selections.

We consider three different linear precoders -- zero-forcing (ZF), maximal ratio transmission (MRT) and minimum mean-squared-error (MMSE) -- for SDMA and RSMA private streams. For the RSMA common stream, we consider the multicast rate maximizing precoder \cite{Hsiao_etal_2015} for $K = 2$ , and the dominant left singular vector of the normalized channel matrix as the precoder for $K = 4$. The power allocation to the RSMA common stream has been optimized using the closed-form suboptimal technique detailed in \cite{mao2022fundmental}, while for both RSMA and SDMA equal power allocation has been assumed for the private streams and across the subcarriers. To capture spatial correlations between the channels of closely located users, we enable the spatial consistency procedure, which is an optional feature of \cite{3gpp_release17}.

\begin{figure*}
    \centering
    \begin{subfigure}{\linewidth}
        \centering
        \includegraphics[width=0.95\linewidth]{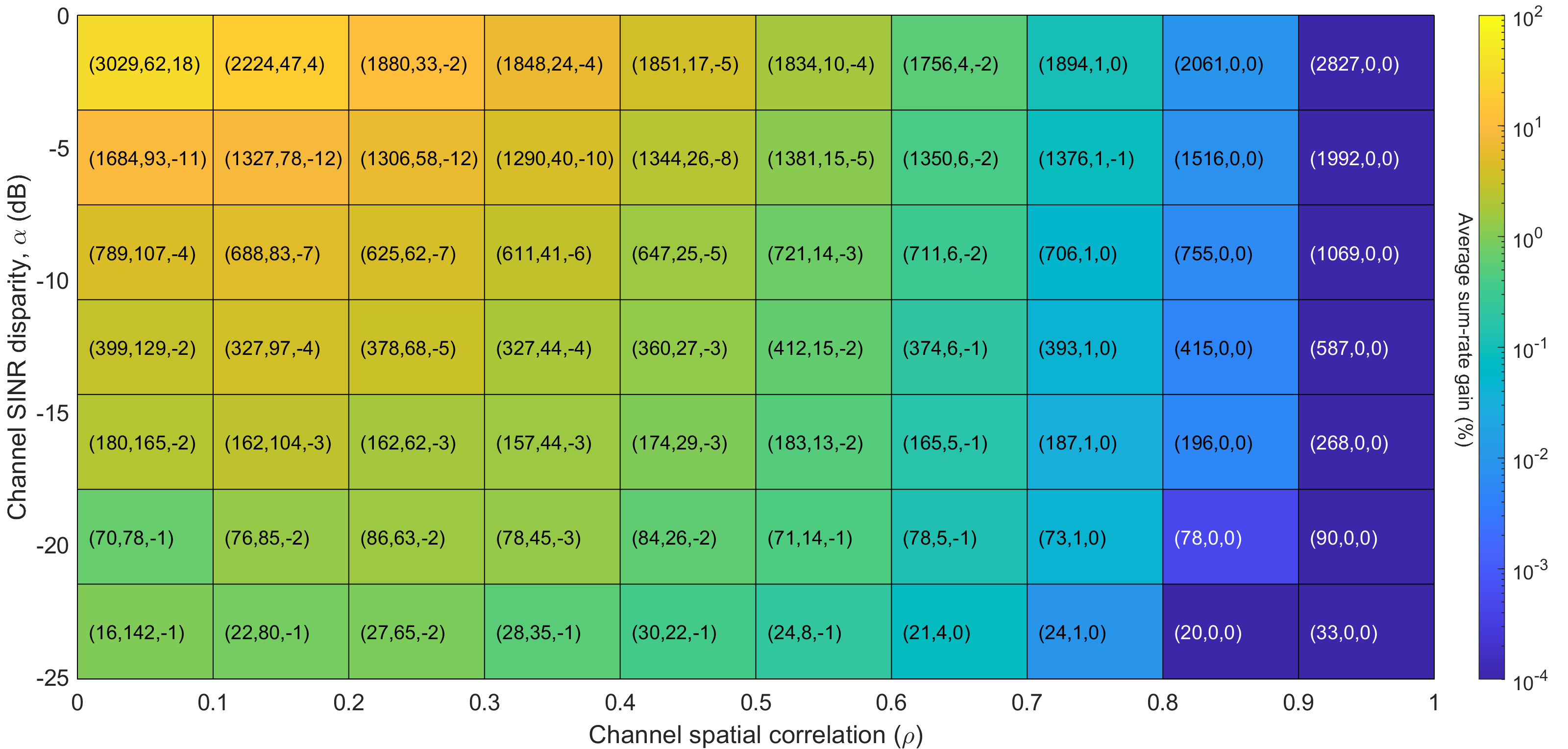}
        \caption{Indoor hotspot (InH) environment}
    \end{subfigure}
    \begin{subfigure}{\linewidth}
        \centering
        \includegraphics[width=0.95\linewidth]{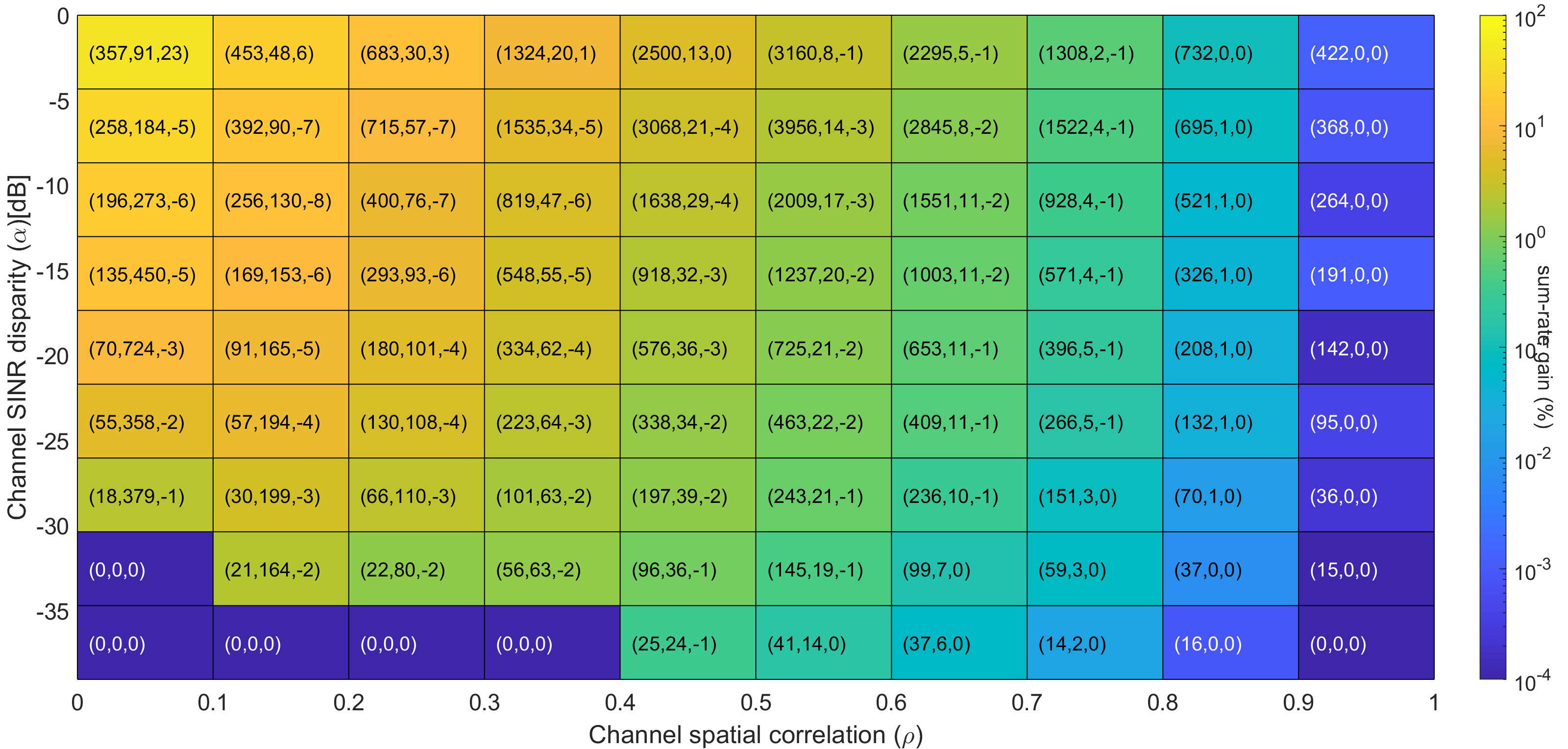}
        \caption{Urban microcell (UMi) environment}
    \end{subfigure}
    \begin{subfigure}{\linewidth}
    \centering
    \begin{tabular}{|c|c|c|c|c|c|}
        \hline
         \multicolumn{2}{|c|}{\% gain in} &  \multicolumn{2}{|c|}{$K = 2$}  & \multicolumn{2}{|c|}{$K = 4$} \\
         \cline{3-6}
          \multicolumn{2}{|c|}{(user rate, weakest user rate) } & 5th percentile & 50th percentile & 5th percentile & 50th percentile \\
            \hline
           UMi  & ZF & (29.0, 26.6) & (10.1, 23.5) & ($>$100, $>$100) & ($>$100, $>$100) \\
                & MRT & (23.1, 23.7) & (32.5, 46.7) & (29.3, 33.9) & (15.9, 44.3) \\
                & MMSE & (19.3, 19.0) & (8.5, 21.7) & (25.8, 34.0) & (9.1, 35.3) \\
            \hline
            InH & ZF & (182.1, 181.7) & (14.6, 34.6) & ($>$100, $>$100) & ($>$100, $>$100) \\
                & MRT & (17.6, 17.8) & (26.0, 36.0) & (10.0, 11.8) & (6.7, 18.6) \\
                & MMSE & (22.6, 20.7) & (7.6, 25.6) & (12.1, 12.5) & (4.1, 15.6) \\
            \hline
    \end{tabular}
    \caption{Percentage gain in (User rate, weakest user rate) for RSMA over SDMA.}
    \end{subfigure}    
    \caption{System level simulations: The colormaps in (a) and (b) present the average percentage sum-rate gains for RSMA over SDMA for $K = 2$ using MMSE precoders. The 50,000 user pairs are binned based on the intervals for the channel spatial correlation ($\rho$) and channel SINR disparity ($\alpha)$ between the users. Each bin also displays the tuple $(N,G_w,G_s)$ where $N$ is the number of data points falling in the bin, and $G_w (G_s)$ is the weaker (stronger) user's average percentage rate gain (obtained by averaging over the $N$ data points). We only consider bins with at least 10 data points; hence, bins where $N < 10$ are displayed as $(0,0,0)$. The table in (c) presents the percentage gains in the user rate for RSMA over SDMA at the 5th and 50th percentile.}
    \label{fig:sls}
\end{figure*}

For $K = 2$, Figs.~\ref{fig:sls}a and b illustrate how the sum rate gains of RSMA over SDMA vary with the channel spatial correlation and the channel signal-to-interference-plus-noise ratio (SINR) disparity. The spatial correlation is measured by the parameter $\rho \in [0,1]$ defined in \cite{RSMA_unicast_prototype}, where $\rho = 0$ signifies fully aligned (high interference) channels while $\rho = 1$ signifies interference-free orthogonal channels. The SINR disparity ($\alpha$), measured in decibels (dB), is defined as the extent to which the weaker user's SINR is lower than the stronger user's SINR -- hence, it is a non-positive quantity. Each bin (characterized by an interval for $\rho$ and $\alpha$) is associated with a tuple $(N,G_w,G_s)$, where $N$ denotes the number of data points falling in the bin, and $G_w (G_s)$ denotes the weaker (stronger) user's average percentage rate gain (obtained by averaging over the $N$ data points). The color of a bin represents the average percentage gain for the sum rate. We observe that:
\begin{itemize}
    \item the sum rate gains are highest towards the upper left hand corner, which represents users that have small SINR disparities (small negative $\alpha$) and high spatial correlations (small $\rho$). These conditions are typically experienced by users in close proximity, which can be frequent in the InH scenario (around 16.5\% -- i.e., 8264 out of 50000 -- of simulation instances satisfy $\rho < 0.2$ and $\alpha  >  -10{\rm dB}$), due to its line-of-sight dominated propagation patterns and relatively small dimensions. Large numbers of closely spaced users are also highly likely in high demand density environments. For UMi on the other hand, users in close proximity are less likely (only 2.9\% -- i.e., 1460 out of 50000 -- of simulation instances satisfy $\rho < 0.2$ and $\alpha  >  -10{\rm dB}$);

    \item RSMA provides better fairness than SDMA, as evidenced by the substantial rate gain for the weaker user (InH: at least 17\% and as high as 165\%, UMi: at least 13\% and as high as 724\%) for $\rho < 0.5$, at the cost of a marginal reduction in the rate for the stronger user (InH: no more than 12\%, UMi: no more than 8\%). 
\end{itemize} 

Fig. 3c shows user-rate gains (\%) and weakest user-rate gain (\%) of RSMA over SDMA, in the UMi and InH scenarios, at the 5th and 50th percentile user spectral efficiencies (i.e., percentile points of the cumulative distribution functions (CDFs) of the individual user rates as defined in \cite{guidelines_itu}). As seen in Fig. 3c, RSMA can provide relevant user-rate gains while providing better user fairness with significant user-rate gains for the weakest user. In particular, RSMA offers large gains ($>$100\%) for $K = 4$ with ZF precoders. This is because with four users, it is more likely that at least two users have high channel spatial correlation in UMi and InH scenarios. Hence, for SDMA, ZF precoders are ineffective at zero-ing out the interference at all users. However, with RSMA, this issue can be mitigated by allocating more power to the common stream, which results in the large gains.

\subsection{Prototypes and Experimental Evaluations}
\label{subsec:prototypes}

\begin{table*}[]
    \centering
    \begin{tabular}{|c|c|c|}
    \hline
     \textbf{Feature} & \textbf{Imperial prototype} \cite{RSMA_unicast_prototype} & \textbf{VIAVI prototype} \cite{VIAVIRSMA}  \\
    \hline     
     Hardware & USRP & USRP + GPU + Server \\
     \hline
     Antennas at BS ($N_T$) & \multicolumn{2}{c|}{$2$} \\
     \hline
     No. of users ($K$) & \multicolumn{2}{c|}{$2, 4$} \\
     \hline
     Bandwidth   & \multicolumn{2}{c|}{$20{\rm MHz}$}  \\
     \hline
     Waveform    & OFDM with IEEE 802.11 specifications & OFDM with 5G NR specifications\\
     \hline
     Precoder design & \begin{tabular}{c}
        Optimization framework \\
        (e.g., maximizing sum rate)   
     \end{tabular} & \begin{tabular}{c}Zero-Forcing, Maximum Ratio Transmission,\\ 
     Singular Value Decomposition \end{tabular}\\ 
     \hline
     Control Signaling & \begin{tabular}{rl}
          \multicolumn{2}{c}{\textbf{IEEE 802.11 based}} \\
          Channel estimation: & LTF (Long-term field) \\
          Demodulation Reference Signal (DM-RS): & LTF
     \end{tabular}& \textbf{5G NR based}\\
     \hline
    MCS levels & Largely based on IEEE 802.11 &  Largely based on 5G NR \\
    \hline
    Channel coding & Polar & LDPC\\
    \hline
    Receiver & SIC & SIC\\
    \hline
    \end{tabular} 
    \caption{Salient features of the RSMA prototypes developed by Imperial College London and VIAVI.}
    \label{tab:prototype_comparison}
\end{table*}

In terms of experimental testbeds, two RSMA prototypes have been independently built using software-defined radios by research groups at Imperial College London \cite{RSMA_unicast_prototype} and VIAVI \cite{VIAVIRSMA}. Both prototypes realize the two-antenna BS ($N_T = 2$) in Fig.~\ref{fig:RSMA_principle}, and Table~\ref{tab:prototype_comparison} compares their distinctive features. Below, we briefly summarize the experimental results obtained from these prototypes that are relevant for two 6G use cases -- eMBb and ISAC.

\subsubsection{eMBb} 
\begin{figure*}
    \centering
    \begin{tabular}{cc}
    \begin{tabular}{c}
    \begin{subfigure}{0.46\linewidth}
        \centering
        \includegraphics[width = \linewidth]{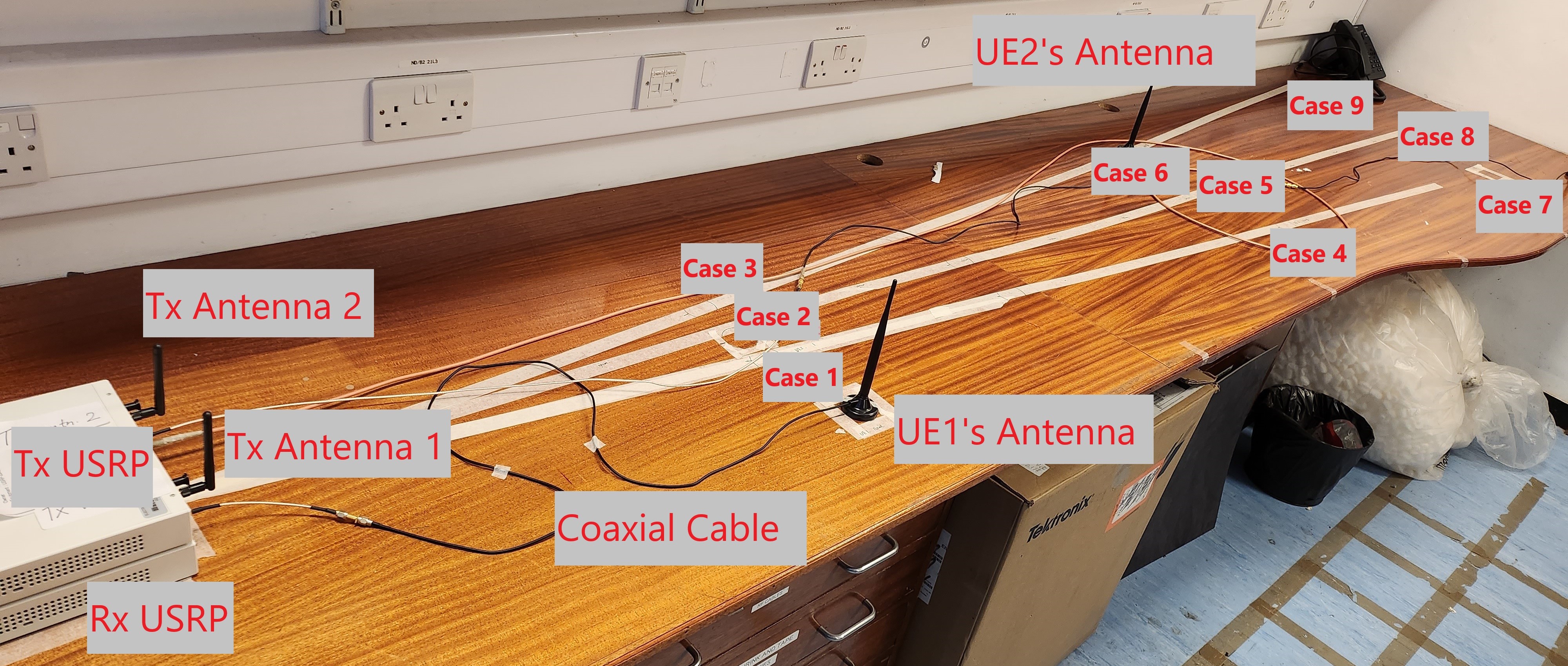}
    \end{subfigure}
    \\
        \begin{subfigure}{0.46\linewidth}
        \centering
        \includegraphics[width = \linewidth]{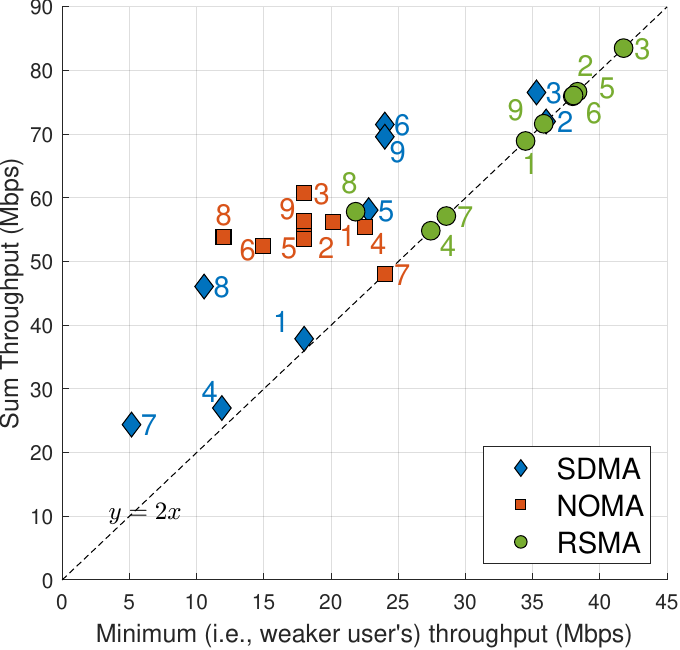}
        \caption{Top: Imperial RSMA prototype and measurement environment. User 1's location is fixed, and we consider nine locations for user 2 to realize pairs of channels capturing variations in channel strength and spatial correlation. Bottom: Three-way RSMA v/s SDMA v/s NOMA comparison of throughput and fairness performance. The numbers 1 through 9 indicate the nine different measurement cases considered. The dashed $y=2x$ line corresponds to max-min throughput fairness. RSMA achieves fairness at a higher sum throughput than both SDMA and NOMA. For more details, see \cite{RSMA_unicast_prototype}.} 
    \end{subfigure}
    \end{tabular}
    &
    \begin{tabular}{c}
    \begin{subfigure}{0.46\linewidth}
        \centering
        \includegraphics[width = \linewidth]{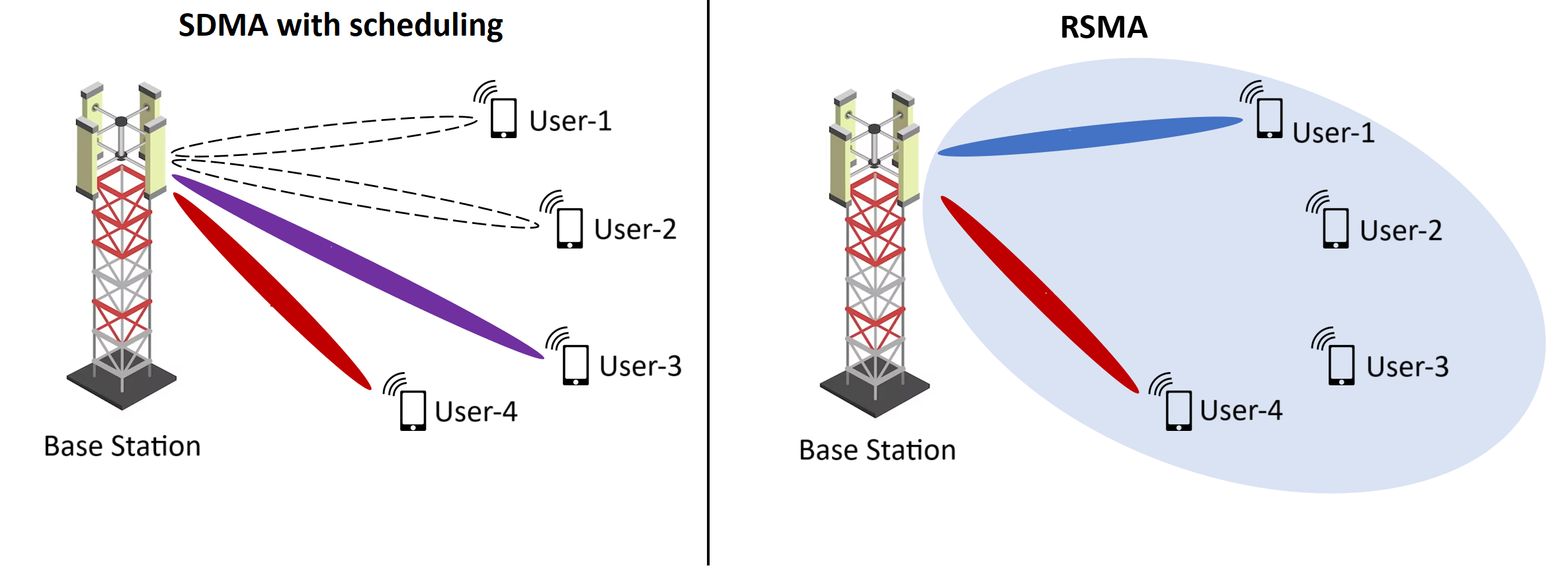}
    \end{subfigure}
    \\
    \begin{subfigure}{0.46\linewidth}
        \centering
        \includegraphics[width = \linewidth]{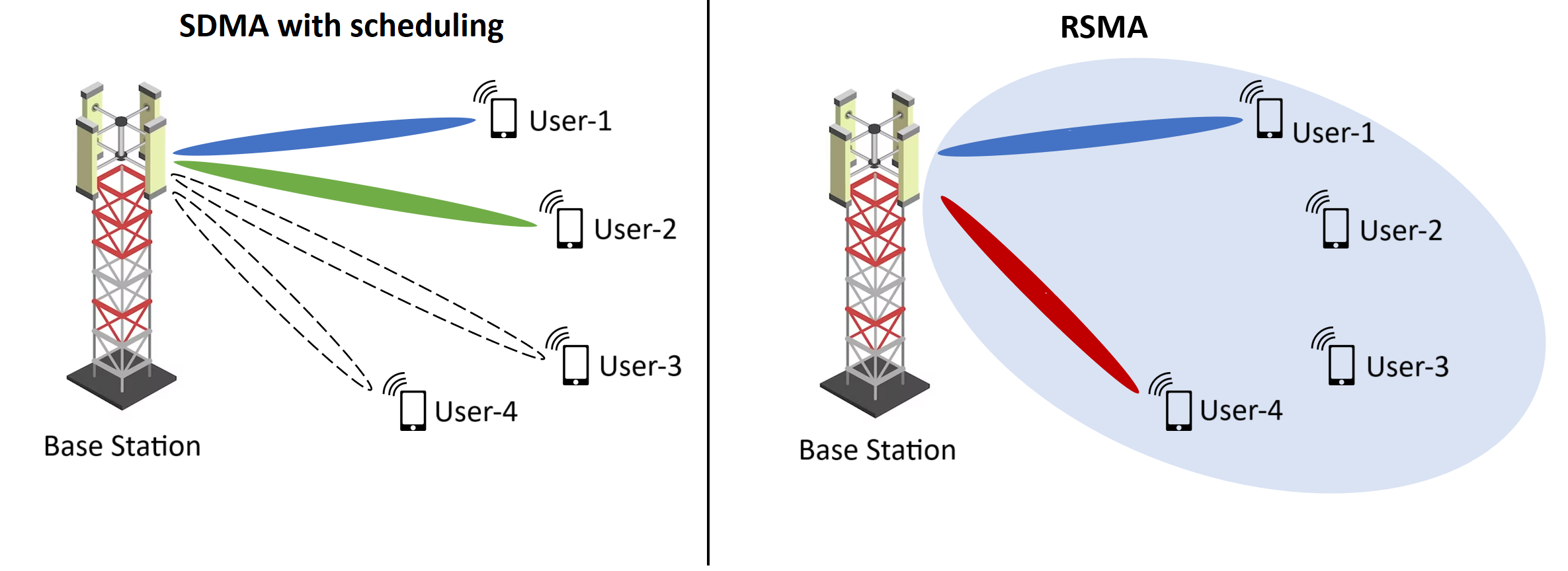}
    \end{subfigure}
    \\
    \begin{subfigure}{0.46\linewidth}
        \centering
        \includegraphics[width = \linewidth]{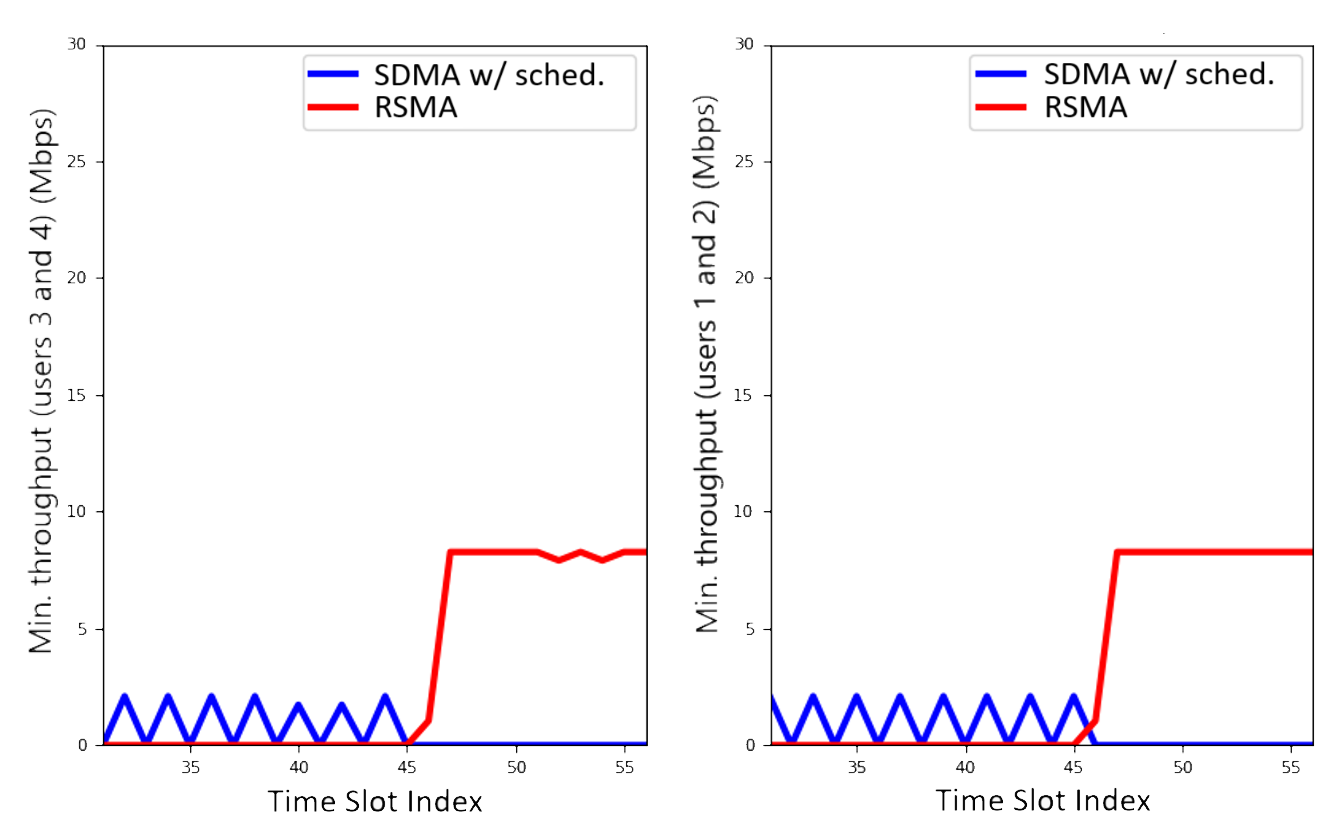}
        \caption{VIAVI RSMA prototype: Slot allocation schemes for RSMA and SDMA+scheduling in overloaded networks -- odd (top) and even (middle) numbered slots. Bottom: Sum-throughput achieved by RSMA and SDMA with scheduling under imperfect CSIT. The throughput performance of SDMA with user scheduling is observed until time slot $45$, after which RSMA is demonstrated to achieve an improved performance without user scheduling.}
    \end{subfigure}
    \end{tabular}
    \end{tabular}
    \caption{RSMA performance for downlink unicast communications, measured using the Imperial and VIAVI prototypes.}
    \label{fig:rsma_experiments}
\end{figure*}

For downlink unicast communications to two users, Fig.~\ref{fig:rsma_experiments}a compares the throughput and fairness performance of RSMA, SDMA and NOMA over nine cases capturing pairs of channels that vary in terms of their relative strength and spatial correlation, measured using the Imperial prototype. Consistent with theoretical predictions, RSMA achieves fairness at a higher sum throughput than both SDMA and NOMA. Fig.~\ref{fig:rsma_experiments}b, on the other hand, compares the max-min rate performance of RSMA in \emph{overloaded} MIMO scenarios -- where the number of transmit antennas is smaller than the number of (single-antenna) users (i.e., $N_T < K$) -- with that of SDMA with user scheduling using the VIAVI prototype. In the overloaded MIMO scenario involving a two-antenna BS and four users, SDMA can serve at most two users in each slot. Hence, scheduling is needed, with users 1 and 2 scheduled in even numbered slots (Fig.~\ref{fig:rsma_experiments}b, middle), and users 3 and 4 in odd numbered slots (Fig.~\ref{fig:rsma_experiments}b, top). However, by allocating two users' messages entirely to the common stream (users 2 and 3), RSMA can serve all users in each slot. The bottom panel of Fig.~\ref{fig:rsma_experiments}b shows how the minimum throughput varies per time slot. SDMA with user scheduling is used to serve the users up to time slot $45$ (the zig-zag throughput pattern reflects the fact that each user is served only in alternate slots), after which RSMA is switched on to serve the same users (the throughput is flat since all users are served in every slot). Clearly, RSMA outperforms SDMA.

% under imperfect CSIT It can be seen from the figure that the minimum rate achieved by the users is improved in the time slots where RSMA is employed. Accordingly, one can conclude that} RSMA outperforms SDMA with user scheduling under imperfect CSIT for certain CSI quality levels. 

%\begin{figure}
%    \centering
%    \begin{subfigure}{\linewidth}
%        \centering
%        \includegraphics[width = \linewidth]{slotA.png}
%        \caption{}
%    \end{subfigure}
%    \begin{subfigure}{\linewidth}
%        \centering
%        \includegraphics[width = \linewidth]{slotB.png}
%        \caption{}
%    \end{subfigure}
%    \begin{subfigure}{\linewidth}
%        \centering
%        \includegraphics[width = \linewidth]{rsma_maxmin.png}
%        \caption{}
%    \end{subfigure}
%    \caption{(a) VIAVI prototype slot allocation schemes for RSMA and SDMA with scheduling - odd numbered slots. (b) VIAVI prototype slot allocation schemes for RSMA and SDMA with scheduling - even numbered slots. (c) Sum-throughput achieved by RSMA and SDMA with scheduling under imperfect CSIT.}
%    \label{fig:rsma_experiments2}
%\end{figure}
Alongside unicast communications, a potentially important variant of eMBb in future networks is \emph{multi-group multicast} (MGM), where each message is desired by more than one user, in general (a group size of one reduces to unicast). Crucially, MGM is a physical layer multicasting technique, amounting to precoder design for one-to-many communications, as opposed to multicasting at higher layers. However, complementary to higher layer multicasting, the motivation for MGM is to realize a more efficient usage of network resources by not imposing a blanket unicast assumption at the physical layer. It is easy to see that MGM applications include live-event streaming, safety-critical vehicular communications, location-based services etc. A distinctive feature of these applications is the high likelihood of encountering an overloaded MIMO scenario. In such scenarios, the interference power at each user cannot be made arbitrarily low through precoding. Thus, the 5G/SDMA strategy of treating interference as noise becomes increasingly ineffective as the number of users increases. Similar to Fig.~\ref{fig:rsma_experiments}a, RSMA achieves fairness at a higher minimum throughput than SDMA and NOMA for an overloaded MGM scenario comprising a two-antenna BS and two groups of two single-antenna users (i.e., $N_T = 2, K = 4$) \cite{RSMA_MGM_prototype}.

Another interesting eMBb use case that combines both unicast and multicast communications is \emph{Non-Orthogonal Unicast Multicast} (NOUM), where a group of $K$ users desire a shared message (multicast), in addition to their respective unique messages (unicast). NOUM applications include live-event broadcasting (shared message: live event stream; unique messages: social media interactions), Vehicle-to-Everything (V2X) networks (shared message: location-dependent information like congestion alerts; unique messages: multimedia entertainment) etc. Essentially, NOUM is a spectrally efficient way to jointly realize physical layer multicast and unicast through linear precoding in multi-antenna systems. Given that the multicast data rate is capped in many NOUM applications (e.g., live-event streaming), RSMA is especially well suited for NOUM, as the common stream can be designed to carry the shared message along with parts of each user's unique message. Just like with unicast communications, the latter components in the common stream help manage unicast interference, which increases the unicast data rate that can be realized while supporting the desired multicast rate \cite{RSMA_NOUM_prototype}.

%\cite{Sibo3GPP} also compares RSMA for joint unicast and multicast services with Orthorgonal/Non-orthorgonal Unicast and Multicast (OUM and NOUM). For multicast service, it is reported that RSMA performs similarly to NOUM, while both are superior to OUM. Meanwhile, RSMA performs better than both NOUM and OUM.

\subsubsection{ISAC}
\begin{figure*}
    \centering
    \begin{subfigure}{0.33\linewidth}
        \centering
        \includegraphics[width=0.7\linewidth]{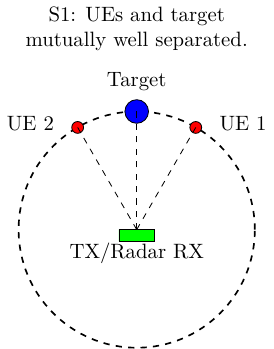}
    \end{subfigure}%
    \begin{subfigure}{0.33\linewidth}
        \centering
        \includegraphics[width=0.8\linewidth]{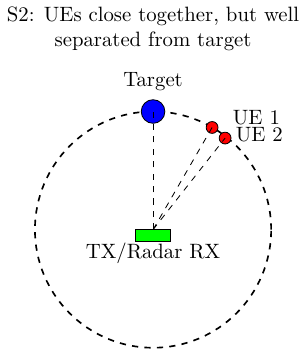}
    \end{subfigure}%
    \begin{subfigure}{0.33\linewidth}
        \centering
        \includegraphics[width=0.63\linewidth]{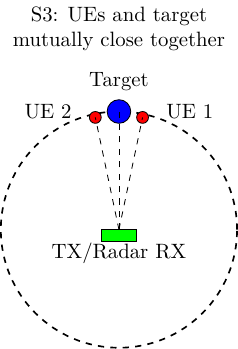}
    \end{subfigure}
    \par \medskip
    \begin{subfigure}{0.33\linewidth}
        \centering
        \includegraphics[width=0.7\linewidth]{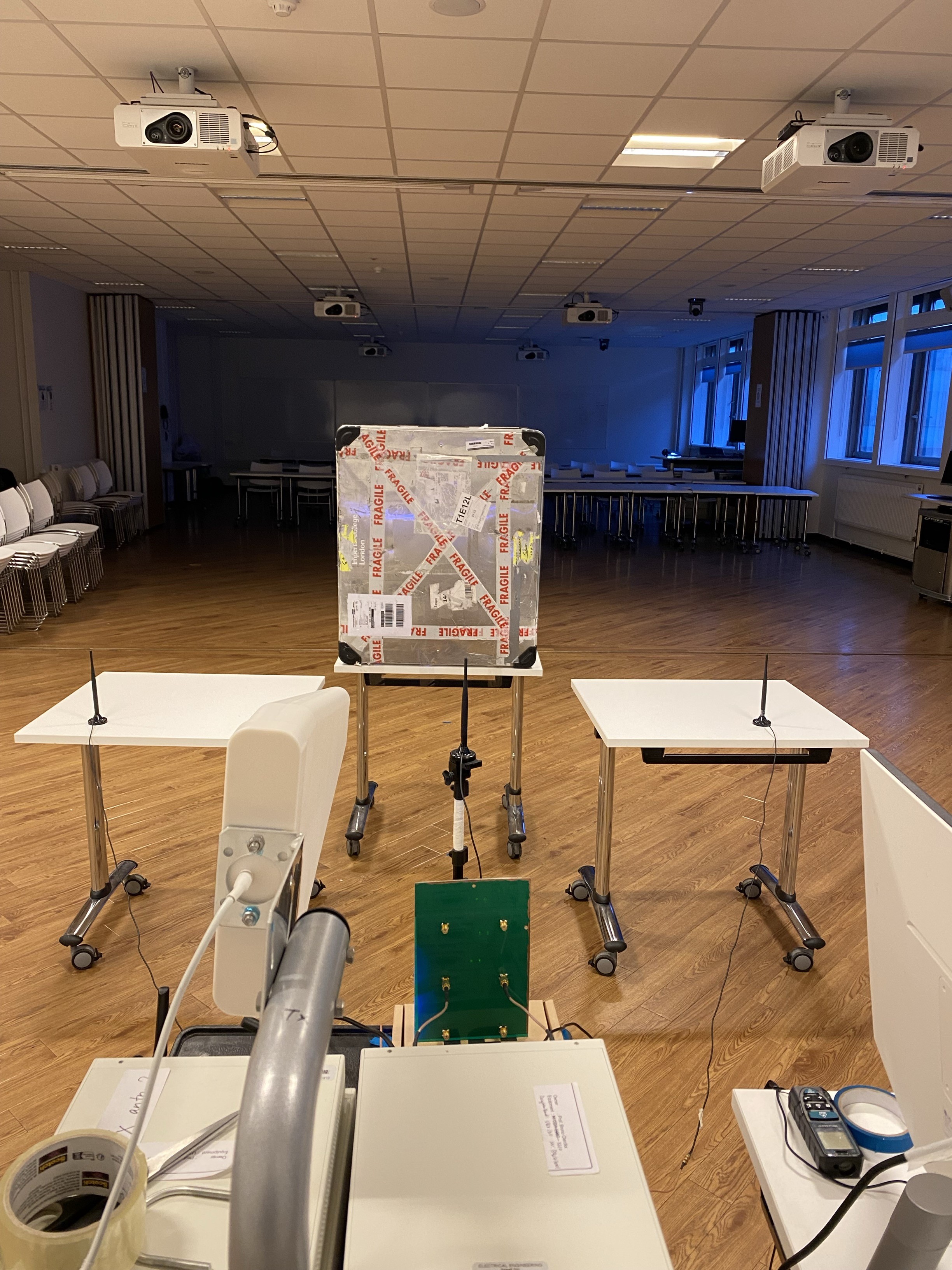}
    \end{subfigure}%
    \begin{subfigure}{0.33\linewidth}
        \centering
        \includegraphics[width=0.7\linewidth]{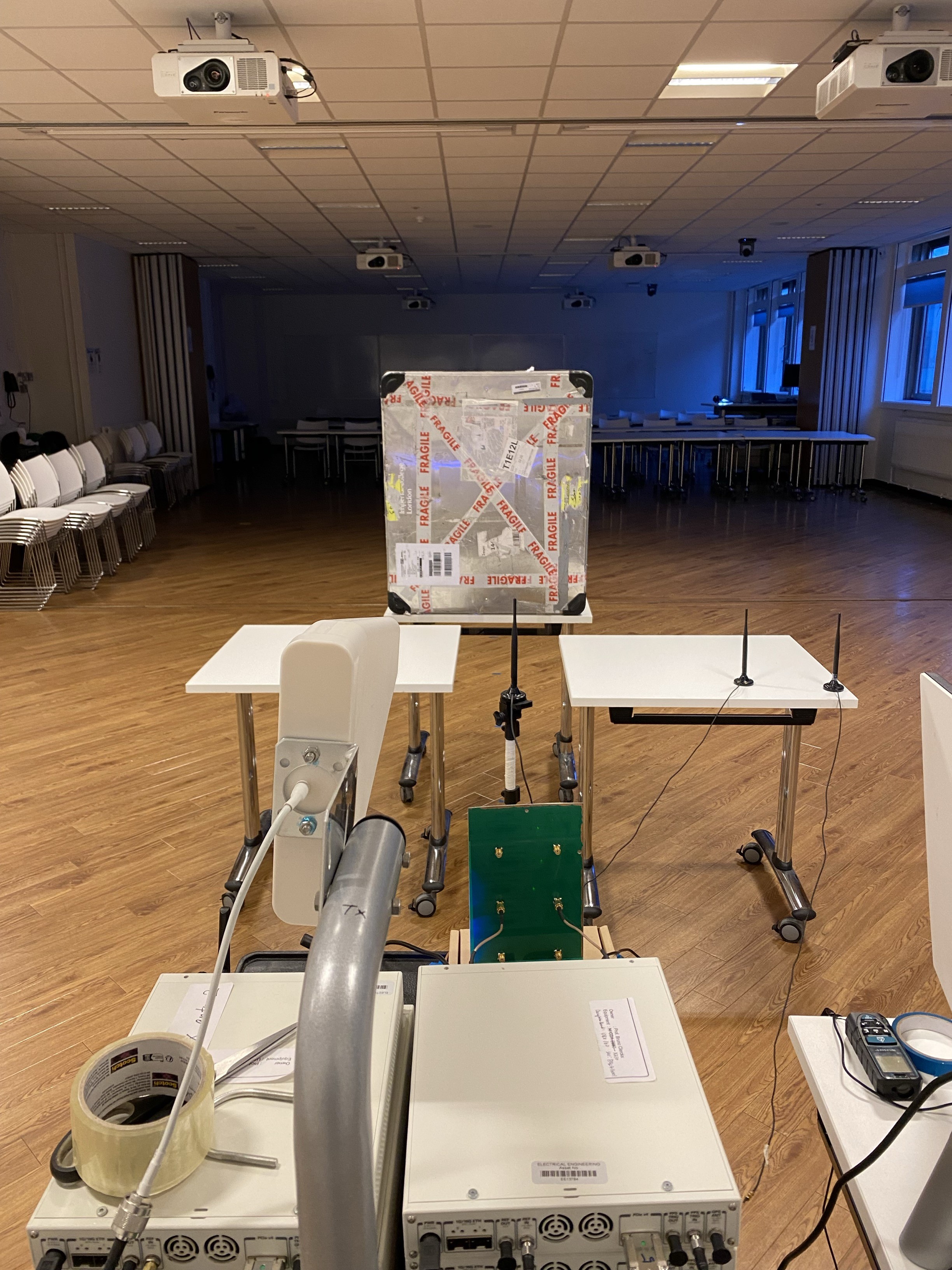}
    \end{subfigure}%
    \begin{subfigure}{0.33\linewidth}
        \centering
        \includegraphics[width=0.7\linewidth]{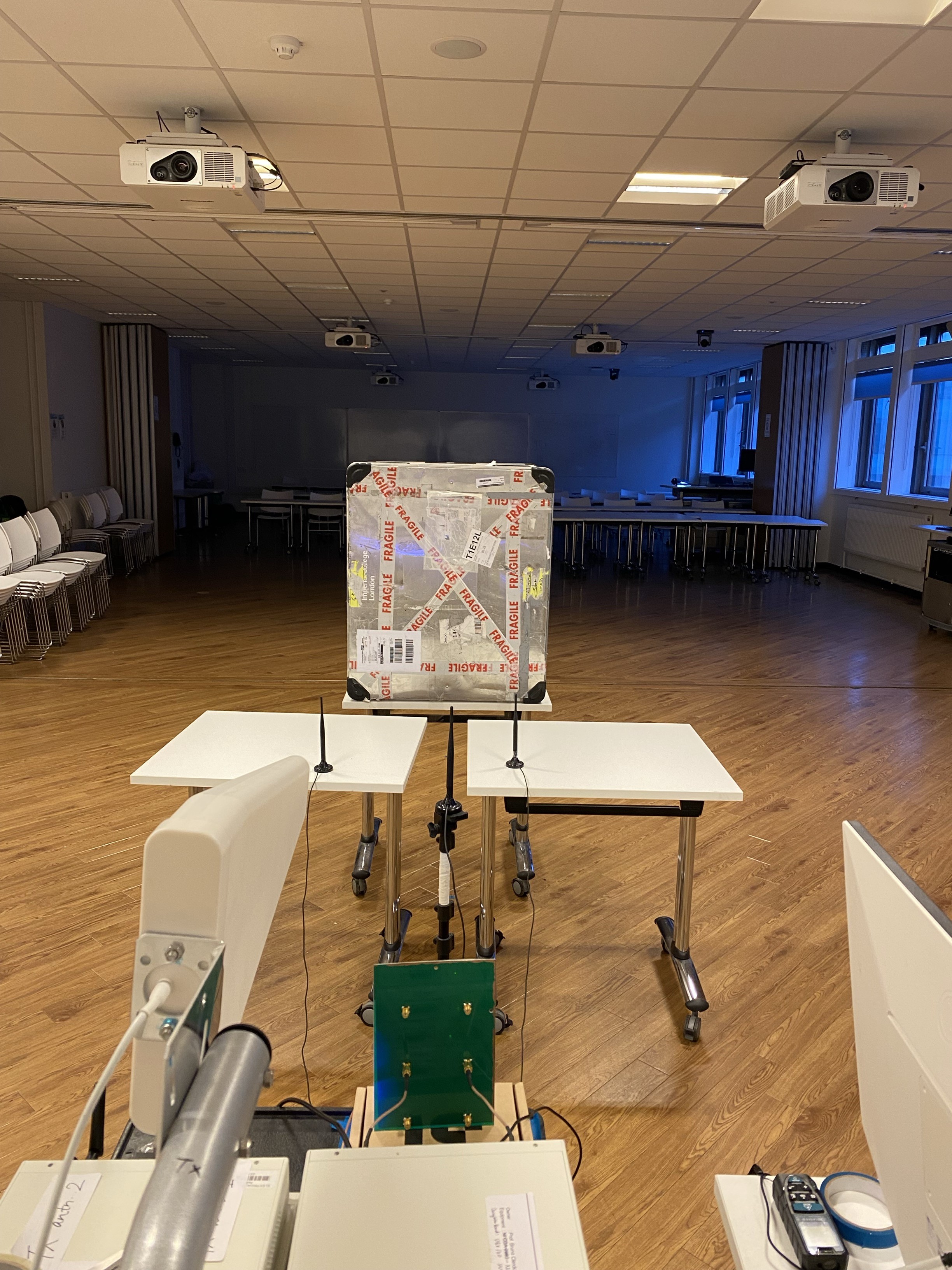}
    \end{subfigure}
    \par \medskip
        \begin{subfigure}{0.33\linewidth}
        \centering
        \includegraphics[width=0.8\linewidth]{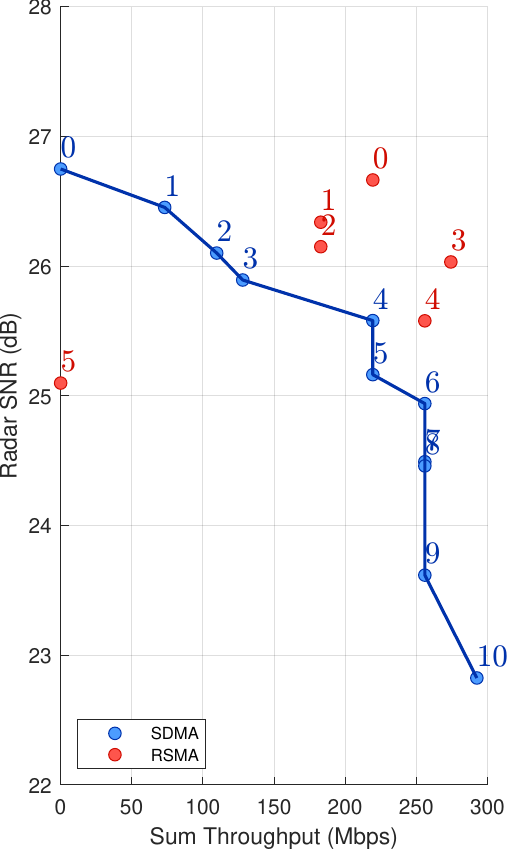}
    \end{subfigure}%
    \begin{subfigure}{0.33\linewidth}
        \centering
        \includegraphics[width=0.8\linewidth]{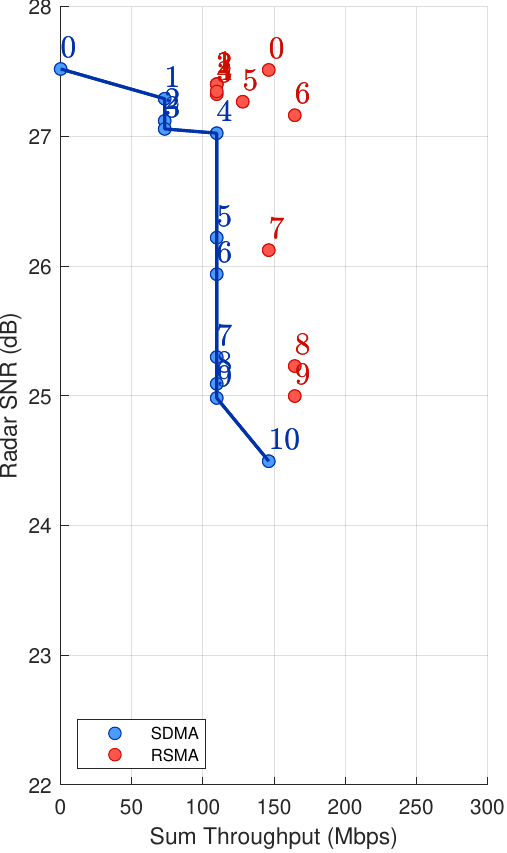}
    \end{subfigure}%
    \begin{subfigure}{0.33\linewidth}
        \centering
        \includegraphics[width=0.8\linewidth]{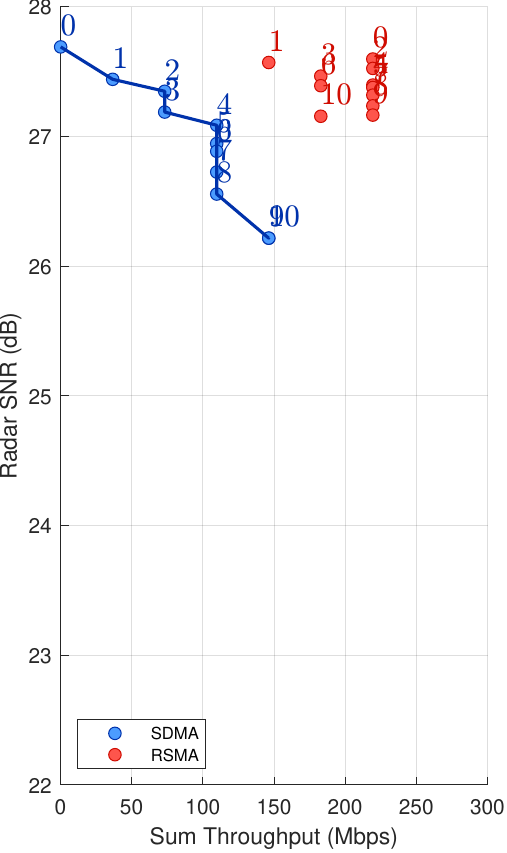}
    \end{subfigure}
    \caption{The ISAC scenarios in the top row capture different levels of inter-user interference and separation/integration between sensing and communications. In our measurements (middle row), the users were approximately $1.5{\rm m}$ away from the TX, while the target was $2.25{\rm m}$ away. These distances, as well as the target dimensions, are representative of peer-to-peer vehicular use cases. The bottom row plots the measured ISAC performance -- RSMA's gain over SDMA is the highest when there is high inter-user interference, as well as high overlap between sensing and communications (in terms of the direction in which power is radiated), as in the rightmost scenario \cite{RSMA_ISAC_prototype}.} 
    \label{fig:rsma_isac}
\end{figure*}

An ISAC use case that is foreseeable for 6G involves a multi-antenna BS simultaneously communicating with $K$ users and sensing targets in its vicinity. As an example, consider $K = 2$ and a single target (Fig.~\ref{fig:rsma_isac}, top row). Precoder design is of utmost importance for this setting, as the limited transmit power budget must be efficiently directed towards both communications users and the target. Several design choices are possible, as listed below: 
\begin{itemize}
    \item[a.] \textbf{Using a dedicated (deterministic) signal for sensing}: For this design choice, SDMA and RSMA have the following features: 
    \begin{itemize}
        \item[i)] \emph{SDMA} involves designing $K+1$ precoders – $K$ for communications and one for sensing. Clearly, the sensing precoder should direct the sensing signal towards the target. On the other hand, since communications signals meant for the users can also be used for sensing \cite{Adi_ccs_ojcoms_2023}, there is a trade-off associated in designing the corresponding precoders - each precoder can be either be directed solely towards the desired user (sacrificing sensing for higher communications performance), or have its radiated power split between the desired user and the target (sacrificing communications performance for better sensing). 

        \item[ii)] \emph{RSMA} involves designing $K+2$ precoders – $K+1$ for communications (from Section~\ref{sec:rsma_overview}), and one for sensing. The remarks on the pointing directions of the communications/sensing precoders from the previous bullet point hold here as well. 
    \end{itemize}

    \item[b.] \textbf{Reusing communications signals for sensing}: In light of \cite{Adi_ccs_ojcoms_2023}, a dedicated sensing signal is arguably an inefficient use of resources, as the power used for the sensing signal offers no communications benefit. Hence, in reusing communications signals for sensing, SDMA and RSMA have the following features:
    \begin{itemize}
        \item[i)] \emph{SDMA} involves designing $K$ precoders, each of which must achieve a balance between radiating power towards its desired user and the target. Thus, each precoder is used for both communications and sensing.

        \item[ii)] \emph{RSMA} involves designing $K+1$ precoders, where the same communications-sensing trade-off from the previous bullet point holds. 
    \end{itemize}
\end{itemize}
Among these choices, RSMA without a dedicated sensing signal (i.e., option b.ii above) yields the largest performance envelope in terms of throughput (communications metric) and radar SNR (sensing metric) \cite{RSMA_ISAC_prototype}, as shown in Fig.~\ref{fig:rsma_isac}. In particular, RSMA's gain over SDMA is the highest when there is high inter-user interference, as well as high overlap between sensing and communications in terms of the direction in which power is radiated (as captured by scenario S3 in Fig.~\ref{fig:rsma_isac}).

\section{Standardization efforts}
\label{sec:stds}
% the 3GPP Organizational Partners -- ARIB, ATIS, CCSA, ETSI, TSDSI, TTA and TTC -- announced that 3GPP will develop the 6th generation of global communications specifications (6G).
\paragraph{3GPP}  At the end of 2023, 3GPP announced its intention to develop the 6G standards. In particular, on Radio Access Network (RAN) aspects, a technical specification group (TSG)-wide 6G workshop will be held in March 2025, while studies on 6G physical layer are expected to start in Release 20 – from the third quarter of 2025 until the first quarter of 2027. Regarding IMT-2030 submission and normative work for 6G in 3GPP, it is expected to start in Release 21.
%Following this announcement, the December 2023 3GPP TSG 102 plenary meetings endorsed in RP-233985 high-level Considerations for 6G Timeline where additional considerations were agreed at 3GPP TSG 103 (RP-240823). 
    
RSMA has neither been studied nor specified in 3GPP for 5G, although it was first proposed in \cite{Sibo3GPP}. It is likely that during the 6G studies there will be discussions on multiple access techniques, where companies will again have the opportunity to propose RSMA as a candidate. 

\paragraph{ETSI Industry Specification Group on Multiple Access Techniques} In December 2024, a new Industry Specification Group (ISG) on Multiple Access Techniques was established at ETSI (called ETSI ISG MAT), which provides an opportunity for ETSI members (and participating non-members) to share their research results and early findings in order to build a wider industry consensus on new multiple access techniques for the upcoming 6G based on 3GPP specifications. The scope of the ISG is on downlink multiple access for the physical layer of the 3GPP radio interface that enhance the transmission efficiency (e.g., spectrum efficiency, power consumption, latency, user fairness, etc.) of specified approaches. Candidate techniques in the scope of the ISG are (but not limited to): orthogonal multiple access, SDMA, NOMA and RSMA.

\section{Conclusion}
The experimental and link/system level simulation results in Section~\ref{sec:rsma_implementation} confirm that RSMA is indeed a promising interference management technique for 6G. It is also a rapidly maturing technology, given the current state of prototyping and standardization activities. However, the main limitations of existing experimental evaluations are:
\begin{itemize}
    \item[i)] the absence of mobility,
    \item[ii)] the small scale in terms of the number of users and antennas at the BS, and 
    \item[iii)] receiver complexity -- with lower complexity non-SIC receiver architectures \cite{SiboTCOM}, (high complexity) SIC is not mandatory at the users to decode the common and private streams.
\end{itemize}
To build momentum towards 6G standardization, it is essential to address the above limitations, while continuing to demonstrate RSMA's gains for more 6G use cases.

\bibliographystyle{IEEEtran}
\bibliography{IEEEabrv, references}
\end{document}